\documentclass{PoS}
\usepackage[intlimits]{amsmath}
\usepackage{amssymb}
\usepackage{graphicx}
\usepackage{mathrsfs,slashed}
\usepackage{epsfig}

\newcommand{\bea}{\begin{eqnarray}}
\newcommand{\eea}{\end{eqnarray}}
\newcommand{\beq}{\begin{equation}}
\newcommand{\eeq}{\end{equation}}

\newcommand{\nn}{\nonumber}

\newcommand{\ii}{{\rm i}}
\newcommand{\ee}{\,{\rm e}}

\DeclareMathOperator{\Tr}{Tr}

\title{NJL model approach to diquarks and baryons in quark matter}

\ShortTitle{NJL model for diquarks and baryons}

\author{D.~Blaschke\\
          Institute for Theoretical Physics, University of Wroclaw, 50-204;
   Wroclaw, Poland \\
   Bogoliubov Laboratory for Theoretical Physics, Joint Institute for
   Nuclear Research, RU-141980 Dubna, Russia\\
       E-mail:  \email{blaschke@ift.uni.wroc.pl}
}

\author{
	 \speaker{A.~Dubinin}
	\\
   Institute for Theoretical Physics, University of Wroclaw, 50-204;
   Wroclaw, Poland \\
   E-mail:  \email{aleksandr.dubinin@ift.uni.wroc.pl}
	    }

\author{D.~Zablocki\\
    Institute for Theoretical Physics, University of Wroclaw, 50-204;
   Wroclaw, Poland \\
       E-mail:  \email{dan.zablocki@gmail.com}
}

\abstract{
	We describe baryons as quark-diquark bound states at finite temperature
	and density within the NJL model for chiral symmetry breaking and restoration
	in quark matter.
	Based on a generalized Beth-Uhlenbeck approach to mesons and diquarks we 
	present in a first step the thermodynamics of  quark-diquark matter which 
	includes the Mott dissociation of diquarks at finite temperature.
	In a second step we solve the Bethe-Salpeter equation for the baryon as a 
	quark-diquark bound state in quark-diquark matter. 
	We obtain a stable, bound baryon even beyond the Mott temperature for diquark
	dissociation since the phase space occupation effect (Pauli blocking for 
	quarks and Bose enhancement for diquarks) in the Bethe-Salpeter kernel for the
	nucleon approximately cancel so that the nucleon mass follows the in-medium
	behaviour of the quark and diquark masses towards chiral restoration.
	In this situation the baryon is obtained as a ``borromean'' three-quark 
	state in medium because the two-particle state (diquark) is unbound while the
	three-particle state (baryon) is bound. 
			}

\FullConference{XXII International Baldin Seminar on High Energy Physics Problems,\\
		15-20 September 2014\\
		JINR, Dubna, Russia}

\begin{document}

\section{Thermodynamics of correlations in quark matter}

While numerous works have recently studied the thermodynamics of quark matter
on the mean-field level including the effects of the medium dependence of the 
order parameters, not so much is known beyond the mean field, about hadronic 
correlations and their backreaction to the structure of the model QCD phase
diagram and its thermodynamics.
Here we will elaborate on the generalized Beth-Uhlenbeck form of the equation
of state which is systematically extended from studying mesonic 
correlations \cite{Hufner:1994ma} to the inclusion of diquark degrees of 
freedom \cite{Blaschke:2013zaa,Blaschke:2014zsa}. 
To that end we will employ a Nambu--Jona-Lasinio-type quark model
with fourpoint interactions in mesonic (quark-antiquark)  and diquark 
(quark-quark) channels.
We shall discuss here the importance of the interplay of the resonant 
states with the residual non-resonant ones  in the continuum of 
scattering states. Due to the Levinson theorem both contributions have the
tendency to compensate each other in quark matter above the Mott transition
\cite{Wergieluk:2012gd}, see also \cite{Blaschke:2013zaa,Blaschke:2014zsa,Yamazaki:2012ux,Rossner:2007ik}.
 
The most intriguing questions will occur when on the basis of this in-medium 
bosonized effective chiral quark model the next step of the hadronization 
program nucleon will be performed and diquarks will be ``integrated out'' in
favor of baryons so that nuclear matter can be described in the model
QCD phase diagram.
A possible scheme for introducing baryons as quark-diquark bound states and 
integrating out the colored and therefore not asymptotically observable 
diquark fields has been suggested by Cahill and collaborators 
\cite{Cahill:1988zi,Cahill:1988bh,Burden:1988dt,Praschifka:1986nf}.
It was afterwards elaborated by Reinhardt \cite{Reinhardt:1989rw} and 
developed further by including the solitonic aspects of a field theoretic 
description of the nucleon \cite{Zuckert:1996nu}.
However, the step to describe nucleonic many-body systems on this quark model 
basis has not been performed by these authors.
Exploratory studies within the framework of an effective local NJL-type model
for the quark-diquark interaction vertex have revealed a first glimpse at the
modification of the nucleon spectral function in the different regions of the 
model QCD phase diagram, including chirally restored and color superconducting
phases \cite{Wang:2010iu}. 
Again, the crucial step towards an equation of state for nuclear matter with 
this microscopic model for nucleons in the QCD phase diagram has not yet been done.
 
We will prepare the ground for a Beth-Uhlenbeck description of nuclear matter,
to be discussed in future work. 
In particular, at zero temperature the structure of a Walecka model for 
nuclear matter shall emerge under specified conditions.
Earlier work in this direction \cite{Bentz:2001vc,Bentz:2002um,Lawley:2006ps}
has demonstrated this possibility although 
no unified description of the nuclear-to-quark matter transition has been 
possible and the elucidation of physical mechanisms for the very transition 
between the hadronic and the quark matter phases of low energy QCD has been 
spared out. Or study aims at indicating directions for filling this gap 
by extending the discussion of the Mott mechanism for the dissociation 
of hadronic bound states of quarks within the NJL model description 
of low-energy QCD which has been given on the example ot two-particle correlations 
(mesons and diquarks) in \cite{Blaschke:2013zaa}
and sketching steps towards a Beth-Uhlenbeck description of nuclear 
matter where nucleons are treated as quark-diquark correlations
in quark matter \cite{Blanquier:2011zz}.

\subsection{Quark matter thermodynamics in the NJL model}

We base the approach on the NJL model Lagrangian including vector and diquark 
interaction channels besides the standard scalar-pseudoscalar chirally 
symmetric interaction for the isospin symmetric case 
($\mu_u=\mu_d=\mu$ and $m_u=m_d=m_0$)

\bea
\mathscr{L}
	&=&
	\bar{q}[\ii\slashed\partial - m_0+\gamma_0\mu]q
	+\mathscr{L}_{\rm int}~,\\
\mathscr{L}_{\rm int}
	&=&
	G_{\rm S}\left[(\bar{q}q)^2+(\bar{q}\ii\gamma_5\tau q)^2\right]
	+ G_{\rm V}(\bar{q}\ii \gamma_\mu q)^2
	+ G_{\rm D}\sum_{A=2,5,7}
	(\bar{q} \ii\gamma_5\tau_2\lambda_Aq^C)
	(\bar{q}^C\ii\gamma_5\tau_2\lambda_A q)~.\nn
\eea

Starting from the Lagrangian, we perform the usual bosonization by means of 
Hubbard-Stratonovich transformations, thus integrating out the quark degrees 
of freedom to obtain a path integral representation of the partition function
(and thus the thermodynamical potential $\Omega$) in terms of composite fields,
mesons ($M = \sigma, \vec{\pi}, \omega_\mu$) 
and (anti-)diquarks ($D, \bar{D}=\Delta_A, \Delta_A^*$, $A=2,5,7$),
\bea
\label{partition}
\mathscr{Z} &=& \int
	\mathcal{D}\sigma\mathcal{D}\vec{\pi}
	\mathcal{D}\omega_\mu
	\mathcal{D}\Delta_A\mathcal{D}\Delta_A^*
	\ee^{-\int{\rm d}^4x_E~
		\left\{
		\frac{\sigma^2+\vec{\pi}^2}{4G_{\rm S}}
		-\frac{\omega_\mu^2}{4G_{\rm V}}
		+\frac{|\Delta_A|^2}{4G_{\rm D}}\right\}
	+\frac{1}{2}\ln\det \{\beta S^{-1}[\sigma,\vec{\pi},\omega_\mu,\Delta_A,\Delta_A^*]\}
		}~,
		\eea
where the inverse propagator is decomposed into mean field and fluctuation, 
$S^{-1}=S_{\rm Q}^{-1} - \Sigma$, with
		\bea
S_Q^{-1} &=& \begin{pmatrix}
		(\ii z_n+{\mu}^*)\gamma_0
		-\boldsymbol{\gamma}\cdot{\bf p}
		-m
		&\Delta_{\rm MF} \ii\gamma_5\tau_2\lambda_2
		\\
		\Delta_{\rm MF}^*\ii\gamma_5\tau_2\lambda_2
		&(\ii z_n-{\mu}^*)\gamma_0
		-\boldsymbol{\gamma}\cdot{\bf p}
		-m
	\end{pmatrix}~,\\
\label{propagator}
\Sigma &=& \begin{pmatrix}
		\sigma +\ii\gamma_5\vec{\tau}\cdot\vec{\pi}
		\hspace{1cm}
		&- \Delta_A \ii\gamma_5\tau_2\lambda_A
		\\
		- \Delta_A^*\ii\gamma_5\tau_2\lambda_A
		& \sigma + \ii\gamma_5\vec{\tau}\,^T\cdot\vec{\pi}
	\end{pmatrix}~.
\label{fluct}
\eea
Here, $z_n=(2n+1)\pi T$ are the fermionic Matsubara frequencies and 
we have introduced the dynamical quark mass $m=m_0+\sigma$ and the 
effective chemical potential $\mu^*=\mu-\omega_{\rm MF}$; 
the vector meson has been redced to the meanfield value of its timelike 
component $\omega_\mu=(\omega_{\rm MF},\boldsymbol{0})$.
In the  present work, we consider NJL case, but we will 
restrict ourselves to the normal phase without color superconductivity 
($\Delta_{MF}=0$).
Moreover, since we want to describe baryons as quark-diquark bound states, 
we have to go beyond Gaussian order in the diquark fields and include an 
infite set of diagrams which can be resummed to define the quark-diquark
Bethe-Salpeter kernel, see \cite{Blanquier:2011zz} and references therein.

As a result, the thermodynamic potential  $\Omega = - (T/V)\ln \mathscr{Z}$  takes the form
\bea
\Omega_{\rm tot}=
	\Omega_{\rm MF}
	+
	\Omega_{\rm M}
	+
	\Omega_{\rm D}
	+
	\Omega_{\rm \bar{D}}
	+
	\Omega_{\rm B}
	+
	\Omega_{\rm res}
	~,
	\label{Omega_tot}
\eea
where the meanfield part $	\Omega_{\rm MF} = \Omega_{\rm cond}+	\Omega_{\rm Q}$ 
consists of a contribution from condensates 
\bea	
	\Omega_{\rm cond}
	=
	\frac{\sigma_{\rm MF}^2}{4G_{\rm S}}
	-\frac{\omega_{\rm MF}^2}{4G_{\rm V}}
	~,
\label{Omega_cond}
\eea
and from dynamical quarks
\bea	
\label{Omega_Q}
\Omega_{\rm Q}
	&=&
	-2N_cN_f\int\frac{d^3p}{(2\pi)^3}\left\{ E_p
	+T\ln\left[1 + {\rm e}^{-(E_p-\mu^*)/T}\right]
	+T \ln\left[1 + {\rm e}^{-(E_p+\mu^*)/T}\right]\right\}~.
\eea
The correlation contributions have the form 
\bea	
	\Omega_{\rm X}
	=
	\pm\frac{d_X}{2}\frac{T}{V}\Tr\ln \left( \beta^{c_X} S_{\rm X}^{-1} \right) ~, 
	~{\rm X=M,~ D,~\bar{D},~B}~,
\label{Omega_X}
\eea
where  for fermions ((B)aryons) holds the plus sign and $c_X=1$ while 
for bosons ((M)esons, (D)iquarks and their antiparticles ($\bar{\rm D}$)) 
holds the minus sign and $c_X=2$.
The coeffficient $d_X$  in (\ref{Omega_X}) is the degree of degeneracy of the state 
$X$ for which the inverse propagator can be given the form \cite{Blaschke:2013zaa}
\bea	
	S_{\rm X}^{-1}
	=\frac{1}{G_{\rm X}} - \Pi_{\rm X}(\ii z_n, {\bf q})
\label{corr-prop}
\eea
with the effective coupling constant $G_X$ being defined by the coupling in the corresponding
interaction channel of the NJL Lagrangian. The coupling for the baryon $G_B$ is discussed below. 
The polarization functions $\Pi_{\rm X}(\ii z_n, {\bf q})$ are defined 
in the RPA approximation as one-loop integrals which involve combinations
of quark-quark, quark-antiquark and quark-diquark propagators with the 
corresponding vertex functions in the meson-, diquark-, and nucleon channels,
respectively. 
Any residual terms which cannot be captured in the above approximation scheme
are summarized in the residual thermodynamic potential $\Omega_{\rm res}$.
Here one finds in particular contributions beyond the Gaussian approximation 
in the meson fields as well as contributions to the baryon-baryon interaction.
The required polarization loop integrals for mesons and diquarks
are given in Ref.~\cite{Blaschke:2013zaa} for the NJL model.
For the Polyakov loop extended NJL model one has to replace in the final expressions
the Fermi- and Bose distribution functions for the color-carrying degrees of freedom 
(Q, D, $\bar{\rm D}$) by the generalized Fermi and Bose distribution functions, respectively 
\cite{Blaschke:2014zsa,Blanquier:2011zz,Hansen:2006ee}.

\subsection{The baryon as quark-diquark state}
\label{ssec:baryon}

The quark-diqark polarization loop, which defines the baryon as an
effective two-particle correlation in the medium, has in general a Dirac structure 
which, however, for a baryon at rest reduces to the scalar function 
(see also  \cite{Blanquier:2011zz}) 
\bea
\label{Pi_B}
\Pi_B(\ii z_n, {\bf 0})&=&4m \int\frac{d^3p}{(2\pi)^3}
\frac{1}{4E_qE_D}\bigg\{
\frac{1-f^-(E_q)+g^-(E_D)}{\ii z_n + E_q + E_D}
-\frac{1-f^+(E_q)+g^+(E_D)}{\ii z_n - E_q - E_D}
\nonumber\\
&&-\frac{f^+(E_q)+g^-(E_D)}{\ii z_n - E_q + E_D}
+\frac{f^-(E_q)+g^+(E_D)}{\ii z_n + E_q - E_D}
\bigg\}~,
\eea 
where $E_q=\sqrt{p^2+m^2}$ and $E_D=\sqrt{p^2+m_D^2}$.
For the PNJL model, the distribution functions for  fermions ($f^\mp(E)=1/[\exp(E\mp\mu)/T +1]$), 
bosons ($g^\mp(E)=1/[\exp(E\mp\mu)/T -1]$) and their antiparticles
need to be replaced by their generalizations for the PNJL case and get then  
strongly suppressed in the confining phase.
\begin{figure}[!h]
\includegraphics[width=0.5\textwidth]{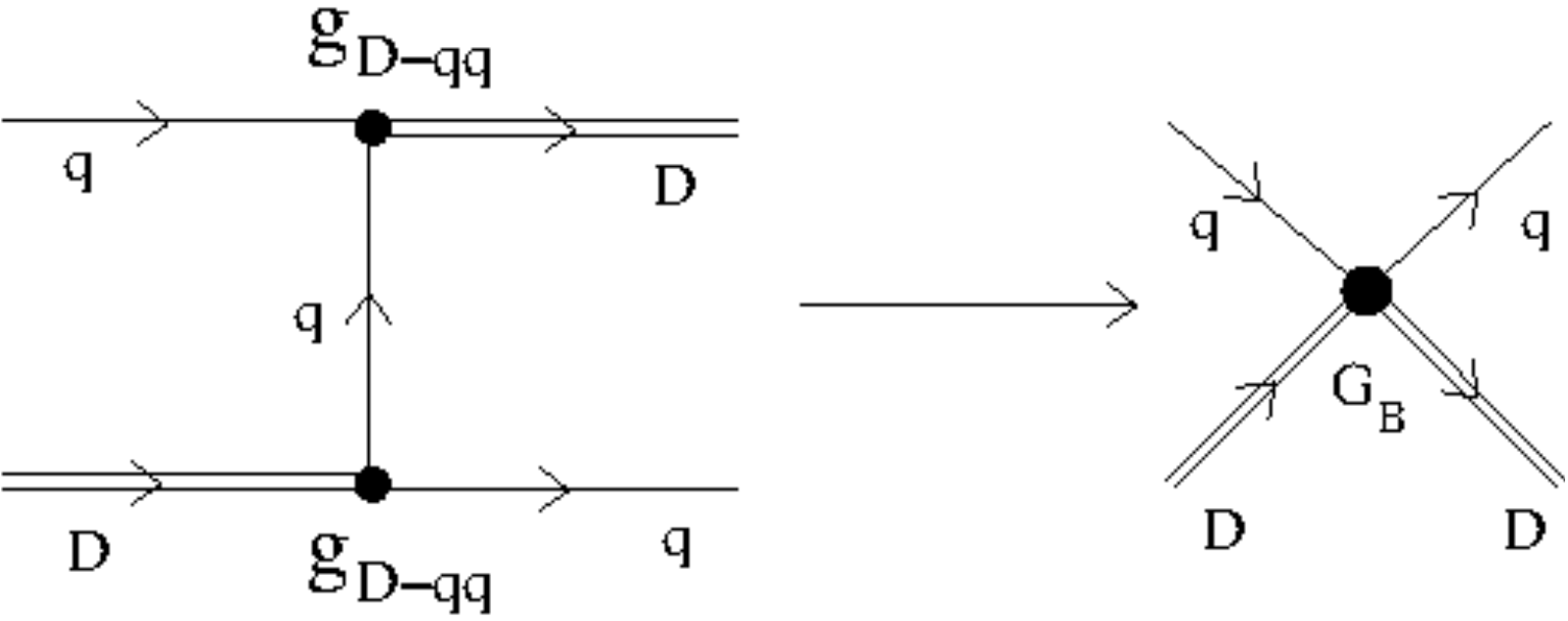}
\caption{Diagrammatic representation of the instantaneous approximation
to the quark-diquark interaction.
\label{fig:vertex}}
\end{figure}

In order to find the baryon mass, we make the instantaneous approximation 
for the quark-diquark interaction vertex (see Fig.~\ref{fig:vertex}) 
which corresponds to identifying 
\bea
G_B=\frac{4 g_{D-qq}^2}{m}~~,~~
g_{D-qq}^2=\frac{4m_D}{\frac{\partial\Pi_D(\omega,{\mathbf 0})}{\partial\omega}
\big|_{\omega=m_D}}\approx {\rm const}~,
\eea
and solve the Bethe-Salpeter equation for the baryon as a bound quark-diquark 
state with a mass $m_B$ according to
\bea
1- G_B \Pi_B(m_B,{\mathbf 0})=0~.
\eea 
Note that in contrast to Ref.~\cite{Wang:2010iu} where $G_B={\rm const}$ was
assumed, we follow here \cite{Blanquier:2011zz} and keep the dependence on the 
quark mass which stems from the instantaneous limit of the quark exchange 
kernel, see Fig.~\ref{fig:vertex}.
Since with increasing temperature towards the chiral restoration the quark mass
drops, this coupling gets enhanced and therefore the tendency for the nucleon 
being a ``borromean state'' which was already noted in \cite{Wang:2010iu} gets
enhanced.

\begin{figure}[!htb]
\includegraphics[width=0.47\textwidth]{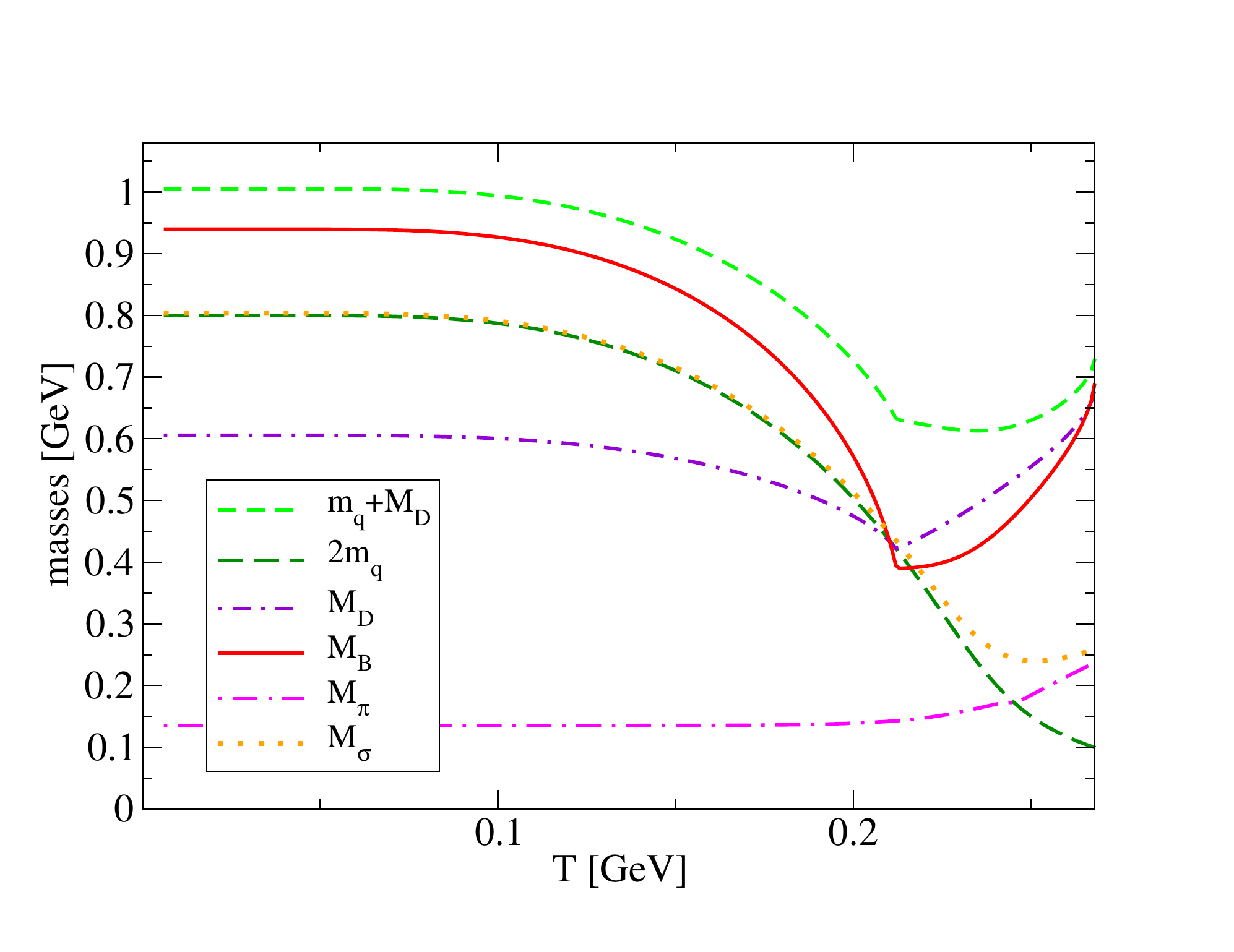}
\includegraphics[width=0.47\textwidth]{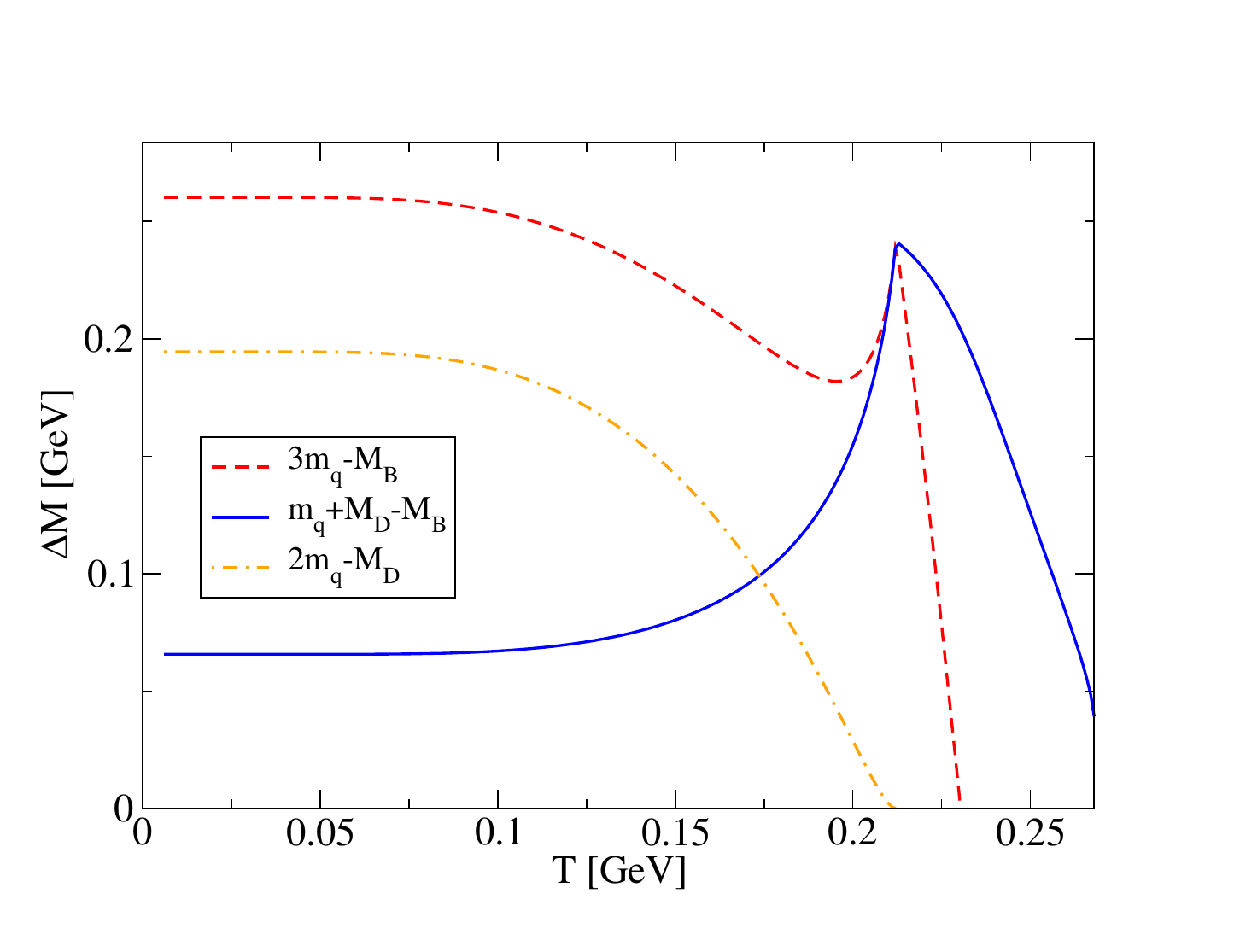}
\caption{Left panel: 
Mass spectrum of pions, sigma mesons, diquarks and baryons as
functions of the temperature $T$. 
Also shown are the relevant thresholds: $2m_q$ for mesons and diquarks and
$m_q+m_D$ for baryons. 
Right panel: Binding energy (mass defect) for baryons relative to the three-quark
or the quark-diquark continuum, resp., and for diquarks relative to the 
two-quark continuum as a function of the temperature $T$. 
The baryon in hot quark matter is a ``borromean'' state: when the diquark 
becomes unbound, the baryon binding energy is still nonvanishing. }
\label{fig:masses1}
\end{figure}

We can solve the mean field gap equation for the quark mass as a function
of temperature and chemical potential for the NJL  model, 
which will serve as inputs for solving the Bethe-Salpeter
equations for mesons, diquarks and baryons in the medium.
The masses of these states are obtained as poles of their propagators and
are shown in Fig.~\ref{fig:masses1}.
We observe that the chiral symmetry restoration which is reflected in the 
dropping quark mass function, induces a Mott effect for the pion and the 
scalar diquark. For both states the kernel of their Bethe-Salpeter equation
contains a Pauli blocking term since they are composed of two fermions.
This Pauli blocking partly compensates the effect of the quark selfenergies
(dropping quark masses) and leads to a stabilization of the bound state
masses against medium effects. This results in the crossing of the 
bound state masses with the continuum threshold, leading to the dissociation
of these bound states. 

Interestingly, the situation is quite different for the baryons.
Here, the bound state is composed of a fermion and a boson, so that there is
a partial compensation of the phase space occupation effects in the Bethe-
Salpeter kernel: the Pauli blocking for quarks gets partly compensated by a
Bose enhancement for the diquarks. 
As a consequence of this, the baryon mass follows the behaviour of the 
quark-diquark continuum and does not experience a Mott effect in the range of
temperatures we studied here, see Fig.~\ref{fig:masses1}.
In the right panel of Fig.~\ref{fig:masses1}, we show the mass defect (binding energy) for 
diquarks and baryons, where for the latter we show two alternative definitions,
i.e., relative to the mass of three quarks and relative to the sum of the 
quark and the diquark mass.
As a remarkable fact we obtain that in a temperature region where the diquark
became already unbound due to the Mott effect, the baryon is still bound.
One can therefore say that the baryon in the vicinity of the chiral 
restoration transition behaves like a ``borromean state'': while the 
two-particle state (diquark) got dissociated, the three-particle state 
(baryon) is still bound. 

\subsection{Thermodynamics of meson, diquark and baryon correlations in quark matter}
\label{ssec:thdyn}
The derivation of the diquark thermodynamic potential follows 
Ref.~\cite{Blaschke:2013zaa} where it was given for the NJL model and the 
form of a generalized Beth-Uhlenbeck EoS was derived.
In such a formulation, the dissociation of bound states in a hot, dense
medium by the Mott effect is given by the behaviour of the in-medium phase 
shifts which encode the analytic properties of the propagator in the complex 
energy plane in a polar representation.
The analoguous result is given in the Ref.~\cite{Blaschke:2014zsa} for the PNJL model. 

Now one is in the state to study the thermodynamics of the meson-, diquark-, and baryon
correlations in a hot and dense medium. 
The equations of state of interest can be derived from the thermodynamical potential 
(\ref{Omega_tot}), e.g., for the pressure $P=-\Omega_{\rm tot}$ or the energy density 
$\varepsilon=\Omega_{\rm tot}-T\partial \Omega_{\rm tot}/\partial T-\mu\partial \Omega_{\rm tot}/\partial \mu$, whereby all corelation contributions 
have the form of a Beth-Uhlenbeck equation 
\bea
\label{GBU_X}
	\Omega_{\rm X}
	=
	\mp d_{\rm X}\int\frac{{\rm d}^3q}{(2\pi)^3}~
	\int_0^\infty\frac{{\rm d}\omega}{2 \pi}~
	\left\{
	\omega
	\pm T\ln Y^{-}_\pm(\omega) \pm T\ln Y^{+}_\pm(\omega) \right\}
	\frac{d\delta_{\rm X}(\omega, {\bf q})}{d\omega}
	~,
\eea
where the upper (lower) sign holds for fermions (bosons)
and $Y^{\mp}_\pm(\omega)=\left(1\pm\ee^{-(\omega\mp \mu_{\rm X})/T}\right)$.
The corresponding degeneracy factors are $d_{\pi}=d_{D}=d_{\vec{D}} =3$ and $d_{B} =4$.
Note that in evaluating 
(\ref{Omega_Q}), (\ref{GBU_X}) 
the zero-point energy terms will be dropped ("no sea" approximation).

For the phase shifts a general decomposition into a resonant (R) and a continuum (c) part can be made
\bea
\label{split}
\delta_{\rm X}(\omega, {\bf q})=
\delta_{\rm X,R}(\omega, {\bf q})+\delta_{\rm X,c}(\omega, {\bf q})~,
\eea
where both parts are uniquely defined by the propagator of the correlation,
see \cite{Blaschke:2013zaa}.
The bound state mass is located at the jump of the phase shift from $0$ to
$\pi$ and this jump corresponds to a delta-function in the Beth-Uhlenbeck 
formulas for the correlations. 
In the case when the continuum of the scattering states  is separated by a sufficient 
energy gap from the bound state, it can be neglected and we obtain as a limiting case the thermodynamics 
of a statistical ensemble of on-shell correlations. 
We shall make use of this fact in the next section when deducing the Walecka model.

\section{The Walecka model as a  low temperature limit of the NJL model}			

In this section, we show that the Walecka model in mean field approximation is 
contained in our model as limiting case.
It is well known, that quarks undergo a Cooper instability leading to diquark 
condensates in this model due to the attractive quark-quark interaction.
At zero temperature, this happens as $\mu^*>m$ or, if diquarks are bound, for 
$\mu^*>m_D/2$.
Here, we restrict our considerations to the region 
$\mu^*< {\rm min} \{m,m_D/2\}$.
However, by construction, the baryon appears as bound state, meaning 
$M_{\rm B}/3< {\rm min} \{m,m_D/2\}$.
Consequently, there is a range in chemical potentials, where the appearance of 
baryons already affects the thermodynamics and leads to non-vanishing baryon 
densities,
\bea
	M_{\rm B}/3<\mu^*< {\rm min} \{m,m_D/2\}~,
\eea
but quarks and diquarks are not excited yet.
One could easily generalize this to finite temperatures, but then the effects
of mesons should be taken into account.

\subsection{Deducing the equations}			

Let us discuss this in more details.
Our thermodynamic potential is given by
\bea
	\Omega_{\rm tot}
	=
	\Omega_{\rm cond}
	+
	\Omega_{\rm Q}
	+
	\Omega_{\rm M,R}
	+
	\Omega_{\rm M,c}
	+
	\Omega_{\rm D,R}
	+
	\Omega_{\rm D,c}
	+
	\Omega_{\rm \bar{D},R}
	+
	\Omega_{\rm \bar{D},c}
	+
	\Omega_{\rm B,R}
	+
	\Omega_{\rm B,c}
	~,
	\label{compoundOmega}
\eea
with the explicit form of each term given above, and where we have decomposed the
meson, diquark and baryon contributions in the respective resonance and continuum parts,
as discussed above.
At zero temperature, there are no contributions from the scattering states, due
to the restriction $\mu^*<m$, which happen to be the thresholds defined in the 
imaginary parts of the meson and diquark polarization functions.
The same holds for the scattering part of the baryons.
This drastically simplifies the calculation as many terms drop out.
However, the low temperature region is already of interest for the properties 
of neutron stars.

We choose the diquark coupling strength $G_{\rm D}$ small enough to avoid the 
diquark pole, and therefore avoid colored objects in this regime, the regime 
where hadrons are the only relevant degrees of freedom. 
Consequently there is no bound state contribution from that part either.
Thus, (\ref{compoundOmega}) reduces to
\bea
	\Omega_{\rm tot}
	=
	\Omega_{\rm cond}
	+
	\Omega_{\rm Q}
	+
	\Omega_{\rm M,R}
	+
	\Omega_{\rm B,R}
	~.
\eea
For the quarks, only the vacuum part contributes in contrast to baryons, due to
the condition $M_{\rm B}/3<\mu^*$.
Explicitly we have
\bea
	\Omega_{\rm tot}
	=
	\frac{\sigma^2}{4G_{\rm S}}
	-
	\frac{\omega_0^2}{4G_{\rm V}}
	-
	2N_cN_f\int_0^\Lambda\frac{{\rm d}^3p}{(2\pi)^3}E_{\bf p}
	-
2N_f\int_0^{\Lambda_{\rm B}}\frac{{\rm d}^3P}{(2\pi)^3}E_{{\rm B},{\bf P}}
	-
	\frac{N_f}{3\pi^2}\int_0^{P_F} {\rm d} P
	\frac{{P}^4}{\sqrt{M_{\rm B}^2 + P^2}}
	~.
\eea
The integrals can be evaluated analytically, 
so that we can write
\bea
	\Omega_{\rm tot}
	=
	\frac{\sigma^2}{4G_{\rm S}}
	-
	\frac{\omega_0^2}{4G_{\rm V}}
	-
	\frac{1}{4\pi^2}
	\left[
		3F(\Lambda,m)
		+
		F(\Lambda_{\rm B},M_{\rm B})
		+
		G(P_F,M_{\rm B})
	\right]
	\label{tdpot1}
	~,
\eea
with functions $F$ and $G$ given in the appendix.
$P_F=\sqrt{(\mu_{\rm B}^*)^2-M_{\rm B}^2}$ is the Fermi momentum of the baryon,
$\mu_{\rm B}^*=3\mu^*$.

\subsection{No sea approximation}			

\begin{figure}[htbp]
\begin{center}
\includegraphics[width=0.49\textwidth]{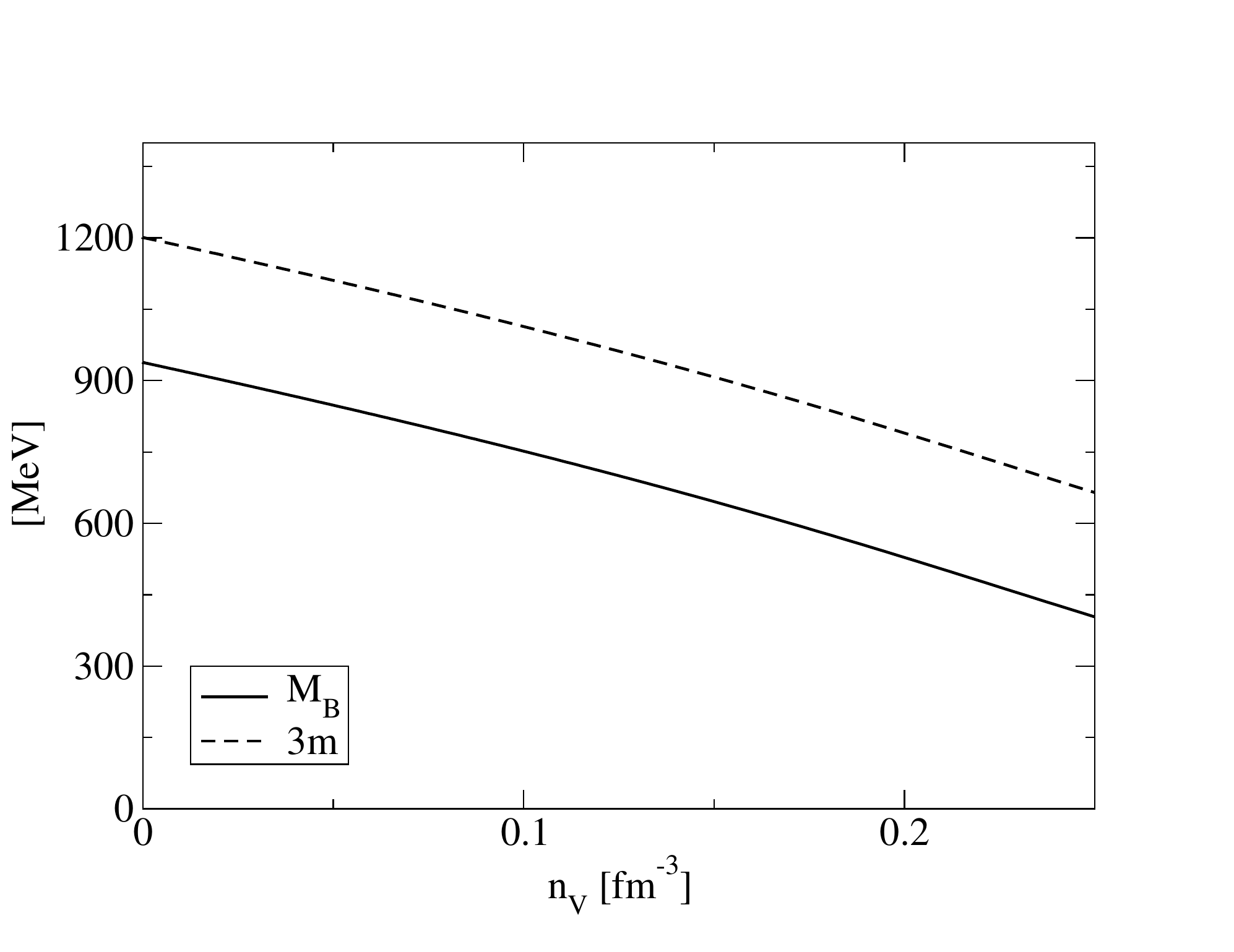}
\caption{
Baryon mass $M_B$ and three-quark continuum ($3m$) as function of the baryon 
density $n_V$.
}
\label{Fig:masses}
\end{center}
\end{figure}

The transition to the original Walecka model needs a few more steps.
First, we drop the vacuum contributions to the thermodynamic potential 
(\ref{tdpot1}); this is what in the Walecka model is referred to as the no-sea 
approximation. The resulting thermodynamic potential reads
\bea
	\Omega_{\rm tot}
	=
	\frac{\sigma^2}{4G_{\rm S}}
	-
	\frac{\omega_0^2}{4G_{\rm V}}
	-
	\frac{1}{4\pi^2}G(P_F,M_{\rm B})
	~.
\label{eq:Om_nosea}
\eea
This already looks like a Walecka type model, but the scalar and vector fields 
here are connected to the quarks.
We now need to clarify the connection between the quark order parameters and 
the corresponding baryonic ones used in the Walecka model.
This is easily achieved by employing quark counting rules, where the vacuum 
contributions are absorbed into the vacuum nucleon mass: 
$M_{\rm B}=3(m_0+\sigma)-\kappa_{\rm B}=M_{\rm B, vac}-3\sigma_{\rm med}$.
Motivated by our finding of Subsection \ref{ssec:baryon} 
(see also Fig.~\ref{Fig:masses}), that the baryon mass defect 
(binding energy) in the model as a quark-diquark bound state is in lowest 
order medium independent, we adopt here in a first step a constant value of 
the parameter $\kappa_{\rm B}=261.7$ MeV.
The solution for the density dependent effective baryon mass is shown in 
Fig.~\ref{Fig:masses}.
The expression for $\sigma_{\rm med}$ proportional to the scalar density 
$n_{\rm S}$ is then obtained from minimizing the thermodynamic potential
(\ref{eq:Om_nosea}) with respect to the order parameter $\sigma$ and setting 
$\Lambda=\Lambda_{\rm B}$,
\bea
	\omega_0
	&=&
	2G_{\rm V}n_{\rm V}
	\\
	n_{\rm V}
	&=&
	\frac{2}{3\pi^2}P_F^3
	\\
	n_{\rm S}
	&=&
	\frac{1}{4\pi^2} F_2(P_F,M_{\rm B})
	=
	\frac{M_{\rm B}}{\pi^2}
	\left[
		P_F E_F-M_{\rm B}^2\ln\frac{P_F+E_F}{M_{\rm B}}
	\right]
	\\
	\sigma_{\rm med}
	&=&
	\frac{3G_{\rm S}}{2\pi^2} F_2(P_F,M_{\rm B})
	=3(2G_{\rm S})n_{\rm S}~.
\eea
We can rewrite the pressure $P=-\Omega_{\rm tot}$ and energy density $\varepsilon$ in terms 
of densities, keeping in mind that the baryon Fermi momentum is 
$P_F=(3\pi^2 n_V/2)^{1/3}$,
\bea
	P
	&=&
	G_{\rm V}n_{\rm V}^2
	-
	9G_{\rm S}n_{\rm S}^2
	+
	\frac{1}{4\pi^2}
	\left(
		P_F E_F\left(\frac2 3P_F^2-M_{\rm B}^2\right)
		+
		M_{\rm B}^4 \ln\frac{P_F+E_F}{M_{\rm B}}
	\right)
	\\
	\epsilon
	&=&
	-P
	+
	\mu n_{\rm V}	
	\\
	&=&
	G_{\rm V}n_{\rm V}^2
	+
	9G_{\rm S}n_{\rm S}^2
	+
	\frac{1}{4\pi^2}
	\left(
		P_F E_F\left(2P_F^2+M_{\rm B}^2\right)
		-
		M_{\rm B}^4 \ln\frac{P_F+E_F}{M_{\rm B}}
	\right).
\eea
Note the factor 9, which appears in front of the scalar coupling strength.
This reminds of the quark origin of this parameter and is in contrast to 
the vector coupling, where there is no prefactor, as the latter is entirely 
due to baryons.
\begin{figure}[!htb]
\begin{center}
\includegraphics[width=0.49\textwidth]{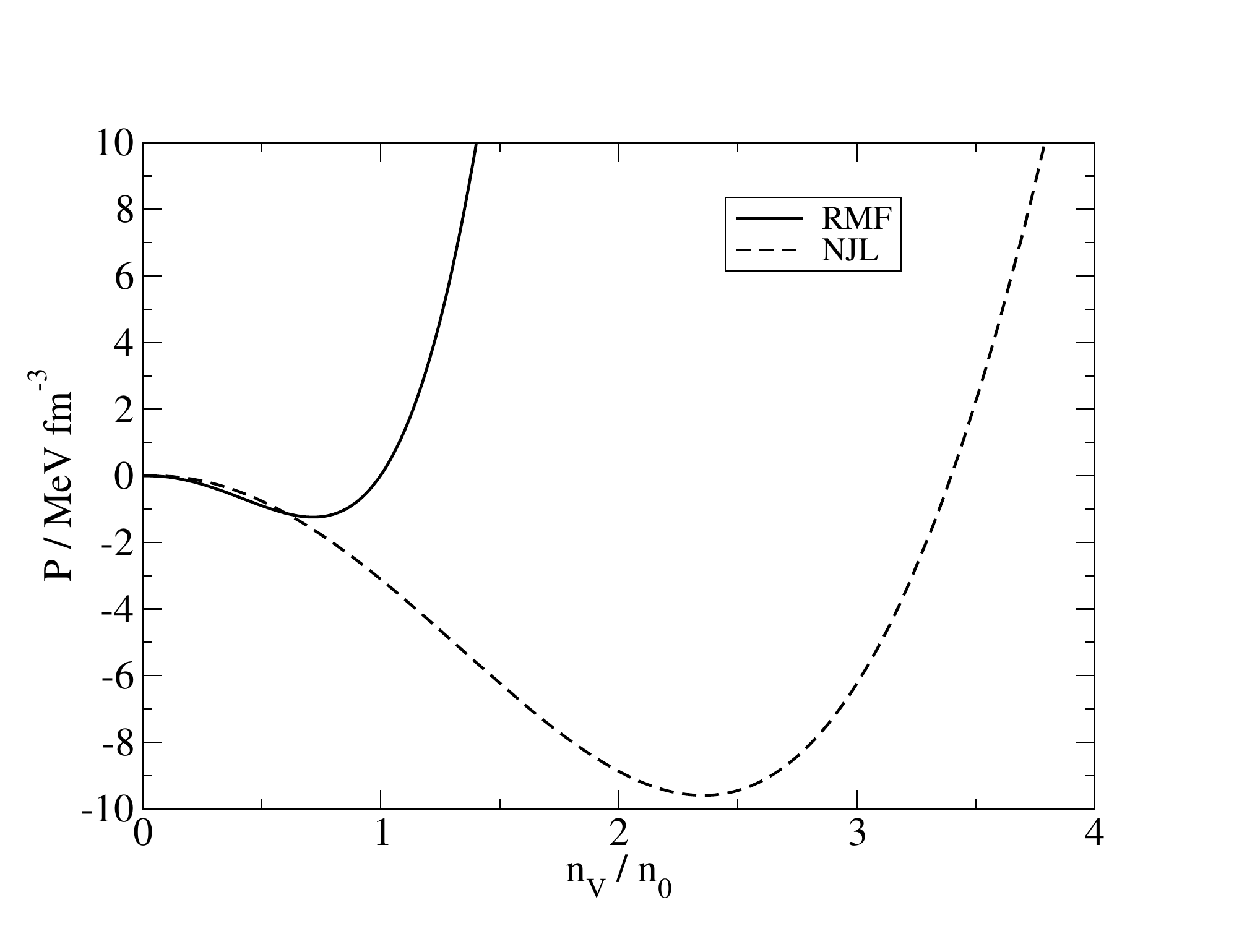}
\includegraphics[width=0.49\textwidth]{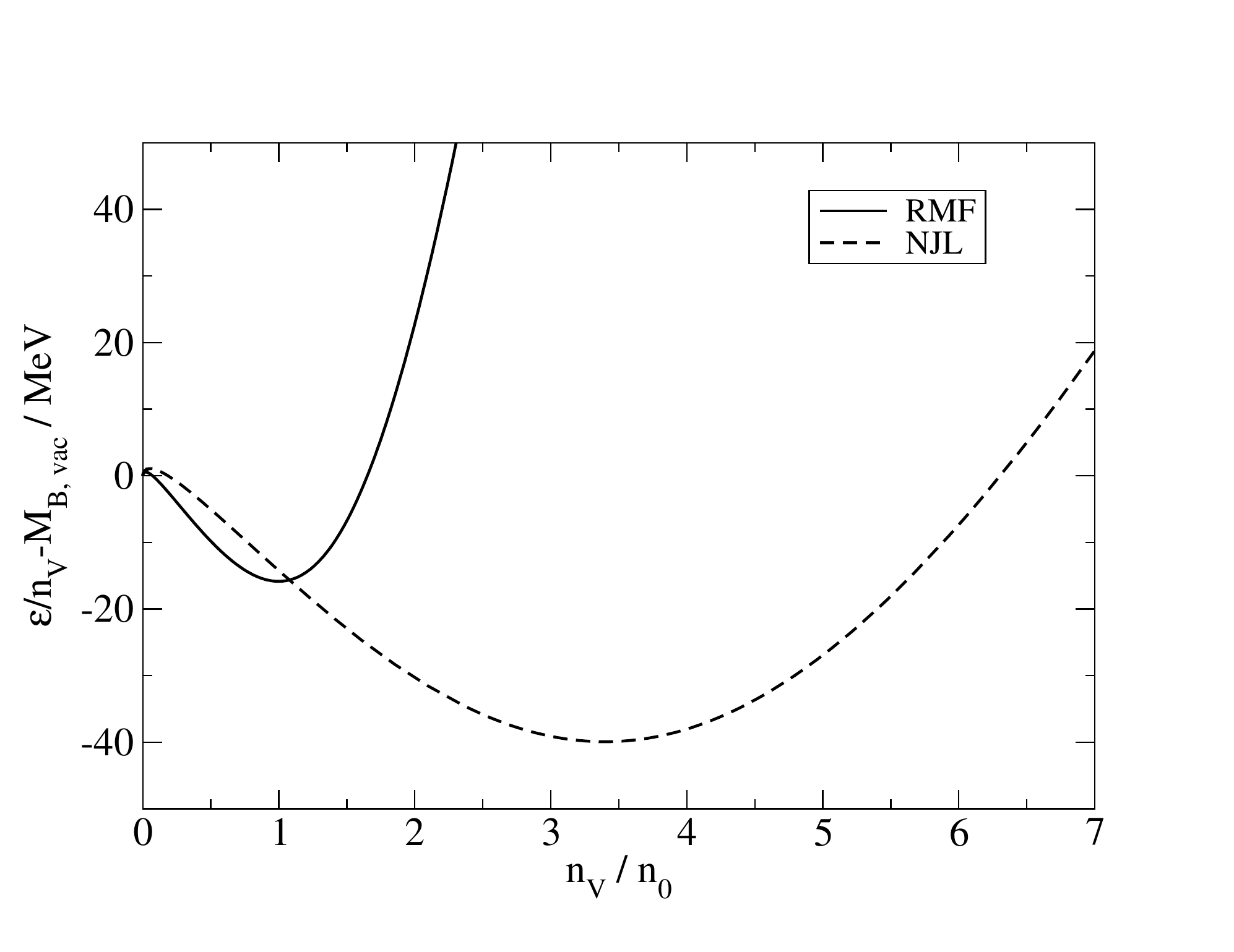}
\caption{
Comparison of the pressure and the binding energy per particle in a fitted 
Walecka model and an unmodified NJL parametrization ($G_V=4.5~G_S$).
For the empirical  value of the saturation density we take 
$n_0=0.16$ fm$^{-3}$.
}
\label{Fig:RMF-NJL-compare}
\end{center}
\end{figure}
We use NJL parameters which would correspond to a constituent quark mass of 
$m=400$ MeV \cite{Grigorian:2006qe}, i.e., 
$m_0=5.588$ MeV, $\Lambda=587.9$ MeV, 
$G_{\rm S}\Lambda^2=2.442$ and $G_{\rm V}=0.5G_{\rm S}$.
Comparing this to the usual expressions for the Walecka model, we can define
NJL model coupling strengths which would reproduce the Walecka model
\bea
	9G_{\rm S, Walecka}&=&\frac{1}{2}\frac{g_\sigma^2}{m_\sigma^2}
	~,~~
	G_{\rm V, Walecka}=\frac{1}{2}\frac{g_\omega^2}{m_\omega^2}~.
\eea
For values, reproducing the binding energy and saturation density 
\cite{Buballa:1996tm}: $m_\sigma=550$ MeV, $m_\omega=783$ MeV, $g_\sigma=10.3$ 
and $g_\omega=12.7$, we can compare with the NJL model parametrization and find instead
\bea
G_{\rm S}&=&0.33~G_{\rm S, Walecka}
	~,~~
G_{\rm V}= 0.024~G_{\rm V, Walecka}~.
\eea
This leads already to saturating matter, even though at too high densities 
($\approx 3.4~n_0$) and with a too large binding energy 
($\approx-40$ MeV instead of $-16$ MeV), 
see Fig.~\ref{Fig:RMF-NJL-compare}.

\section{Discussion and outlook}			

The general aim of our work is the derivation of the thermodynamics for a meson-baryon
system (quantum hadrodynamics = QHD) on the basis of quark degrees of freedom and to
explore the application limits of QHD, e.g., in the case of relativistic mean-field models for 
nuclear matter in heavy-ion collisions and neutron stars.
The present contribution should be seen as a first step in this direction.

We could show that, when the analysis is restricted to a mean-field approximation for mesons
and to small densities below the quark threshold ($\mu_B < 3m_q$), the structure of the 
well-known $\sigma-\omega$ model (Walecka model) for nuclear matter emerges in the limit
$T\to 0$.
In this way we obtain from the NJL model an equation of state for nuclear matter at finite densities and
temperatures, which has a saturation property.
Taking the model as it is and using the standard model parameters, however, we could not 
reproduce the phenomenological values for the saturation properties.
In fact, taking into account the simplicity of the model, this was not to be expected.
Nevertheless, it should be viewed as an achievement of this work that  a general path from 
a chiral quark model to the thermodynamics of nuclear matter could be identified.

In the present case, we find that the saturation density and the binding energy of the 
NJL nuclear matter, which result from the interplay between the scalar and vector mean fields,
are too large.
In order to obtain the phenomenological values, one should take in to account additional effects
which lead to an enhancement of both,  the repulsion (vector field) and of the attraction
(scalar field). 
Here we think that the consideration of higher-order terms in the meson fields could be
a promising option,
because a consistent bosonization of the NJL model requires the consideration and
renormalization of terms up to the fourth order, which would otherwise lead to divergencies.
In this way, one obtains, for example, from the NJL model the successful hadronic
linear $\sigma$ model, see \cite{Klevansky:1992qe} and references therein.
Alternatively, terms of higher order can already be introduced on the level of the original quark 
Lagrangian, leading, however, to new coupling constants 
(see \cite{Benic:2014iaa} and refs. therein). 
This freedom has been used, e.g., by Huguet et al. \cite{Huguet:2007uh} to  
successfully reproduce the saturation properties of nuclear matter  at $T=0$ in this way. 
Other interesting approaches to describe nuclear matter based on quark degrees of freedom
are discussed in \cite{Blaschke:2013zaa}.
Besides this, we have omitted terms in $\Omega_{\rm res}$ which correpond to quark exchange
terms between baryons and would result in a repulsive Pauli blocking contribution at high densities.
Considering such higher order terms one should eventually be able to reduce the binding energy and saturation 
density towards their physical values!

Finally, we want to point out that a suitable confining mechanism has to be present
in a quark-matter model, in order to exclude the appearance of unphysical degrees of freedom
already in the experimentally well studied nuclear-matter regime.
In this context the Polyakov loop has proven to be an interesting tool, although it does not 
involve all aspects of confinement. This is in particular the case at low temperatures and 
for $T=0$.

\section{Acknowledgement}			

We are grateful to M. Buballa for his continued interest in the progress of this work
and his critical remarks which helped improving this manuscript.
This work was supported by the Polish National
Science Centre (NCN) under contract number
UMO-2011/02/A/ST2/00306 (A.D., D.B., D.Z.)
and by the German Deutsche Forschungsgemeinschaft (DFG)
under contract number BU 2406/1-1 (D.Z.). 
A.D. acknowledges a grant from the Institute for Theoretical
Physics of the University of Wroclaw under contract No. 1356/M/IFT/13 
and support by the Bogoliubov-Infeld programme for scientific 
collaboration between Polish Institutions and the JINR Dubna.

\appendix			
\renewcommand*{\thesection}{\Alph{section}}
\section{Basic integrals}
\label{app:integrals}	
We list some integrals encountered in the zero temperature evaluation of quark 
loop integrals
\bea
	 F(x,y)&=&8\int^x{{\rm d}p\ p^2}~\sqrt{p^2+y^2}
	=
	\left(
		x\sqrt{x^2+y^2} (2x^2+y^2) - y^4 \ln\frac{x+\sqrt{x^2+y^2}}{y}
	\right)
	\\
	 G(x,y)&=&\frac{8}{3}\int^x{{\rm d}p}~\frac{p^4}{\sqrt{p^2+y^2}}
	=
	\left(
	x\sqrt{x^2+y^2}\left(\frac2 3x^2-y^2\right)+y^4 \ln\frac{x+\sqrt{x^2+y^2}}{y}
	\right)
\eea
Partial derivatives of these functions are encountered in the evaluation of 
generalized gap equations; here we require
\bea
	F_2(x,y)\equiv
	\frac{\partial F(x,y)}{\partial y}
	=
	4y\left[x\sqrt{x^2+y^2}-y^2\ln\frac{x +\sqrt{x^2+y^2}}{y}\right]
	~.
\eea


\end{document}